\documentclass[12pt,preprint]{aastex}

\shorttitle{Eta Carinae Binarity}
\shortauthors{Soker}
\def \kev{\rm{keV}}
\def \ev{\rm{ev}}

\def \cm{~\rm{cm}}
\def \s{~\rm{s}}
\def \km{~\rm{km}}

\def \K{~\rm{K}}
\def \g{~\rm{g}}

\def \AU{~\rm{AU}}
\def \erg{~\rm{erg}}

\def \yr{~\rm{yr}}

\def \etc{$\eta$~Car}

\begin{document}

\title{ACCRETION ONTO THE COMPANION OF ETA CARINAE DURING THE SPECTROSCOPIC EVENT:
\newline
III. THE He II $\lambda$4686 LINE}

\author{Noam Soker and Ehud Behar \altaffilmark{}}
%\altaffilmark{}
\affil{Department of Physics, Technion$-$Israel
Institute of Technology, Haifa 32000 Israel;
soker@physics.technion.ac.il; behar@physics.technion.ac.il.}

\begin{abstract}

We continue to explore the accretion model of the massive binary
system \etc\ by studying the anomalously high He~II~$\lambda$4686
line.  The line appears just before periastron and
disappears immediately thereafter. Based on the He~II
$\lambda$4686 line emission from O-stars and their modeling in
the literature, we postulate that the He~II $\lambda$4686 line
comes from the acceleration zone of the secondary stellar wind.
We attribute the large increase in the line intensity to a slight
increase in the density of the secondary stellar wind in its
acceleration zone.
The increase in density could be due to the ionization and subsequent
deceleration of the wind by the enhanced X-ray emission arising from
the shocked secondary wind further downstream or to accretion of the
primary stellar wind.
Accretion around the secondary equatorial plane
gives rise to collimation of the secondary wind, which increases
its density, hence enhancing the He~II $\lambda$4686 emission line.
In contrast with previous explanations, the presently proposed model
does not require a prohibitively high X-ray flux to
directly photoionize the He.

\end{abstract}
% ==========================================================
\section{INTRODUCTION}
\label{sec:intro}
% ==========================================================

The massive stellar binary system $\eta$ Carinae displays a
periodicity of $5.54$ years in many emission lines and in the
continuum, from the IR (e.g., Whitelock et al.\ 2004) to the X-ray
(Corcoran et al.\ 2001a; Corcoran 2005; Corcoran et al.\ 2004a,b).
The fading of high excitation lines {{{ (e.g.,
Fe~III~$\lambda$1895, Fe~III~$\lambda$4701,
Ne~III~$\lambda$1747-54, Si~III~$\lambda$1892, Zanella et al.
1984; He~I~$\lambda$10830, Damineli 1996; He~I~$\lambda$6678,
Damineli et al. 2000; and many more lines listed by Damineli et
al. 1998) }}} every $5.54$~years is assumed to occur near
periastron passages, and it is termed the spectroscopic event
(e.g., Damineli et al.\ 2000). The primary star of the \etc\
binary system is more massive and cooler than the hotter secondary
companion, probably an O-type or a WR star (Iping et al. 2005).
The primary wind is slower but with a higher mass loss rate than
the secondary wind. Although it is widely accepted that $\eta$ Car
is a binary system (e.g. Damineli 1996; Damineli et al.\ 1997,
2000; Ishibashi et al.\ 1999; Corcoran et al.\ 2001a,b; 2004b;
Pittard \& Corcoran 2002; Duncan \& White 2003;, Fernandez Lajus
et al.\ 2003; Smith et al.\ 2004; Whitelock et al.\ 2004; Verner
et al.\ 2005), the precise influence of the secondary on the
outflowing gas, and in particular its role in shaping the bipolar
nebula$-$the Homunculus$-$is only poorly understood. The $\ga 10
M_\odot$ Homunculus (Smith et al.\ 2003) was formed during the
1837-1856 Great Eruption, and is now expanding outward at a
velocity of $\sim 600 \km \s^{-1}$ (Davidson \& Humphreys 1997). A
weaker bipolar eruption occurred during the Lesser Eruption of
1890 (Ishibashi et al.\ 2003).

The similarity of the Homunculus morphology to the morphologies
of many planetary nebulae and symbiotic nebulae suggests
that during the Great Eruption the secondary accreted mass
from the primary along the entire orbit and blew two jets, which shaped the
Homunculus (Soker 2001; 2004).
Unlike the case during the 19th century Great Eruption, presently the
companion does not accrete mass during most of its orbital motion.
However, recently Soker (2005a,b) has suggested that for several
weeks near periastron passages the secondary does accrete mass from the primary.
This accretion shuts down the secondary's wind, hence the X-ray emission,
leading to the long X-ray minimum (Soker 2005a,b; Akashi et al.\ 2006).
The formation of an accretion disk in the present day $\eta$ Car binary system
(albeit only for a short time; Soker 2003, 2005a) is in agreement with some of the
suggestions made earlier by van Genderen et al.\ (1994, 1995, 1999).

According to the accretion model,
the basic accretion process near periastron passages is as follows (Soker 2005b):
The collision between the two winds creates (two) shock waves that heat the winds
to X-ray temperatures, which later escape the system.
The post-shock primary wind material cools to a very low temperature, $\la 10^4 \K$,
and is prone to instabilities that might lead to the formation of dense blobs.
While during most of the 5.54~years orbital period the
secondary's gravity has a negligible effect on the winds,
near periastron the collision region is close to the secondary star, and the
secondary's gravitational field becomes significant,
to the extent that the primary wind falls onto the secondary, which in turn
accretes primary wind fragments.
Namely, near periastron passage the collision region of the two winds
is assumed to collapse on the secondary star, a process that
is further assumed to shut down the secondary wind (Soker 2005b).
This substantially reduces the X-ray luminosity (Akashi et al.\ 2006).
The accretion process may last somewhat longer than the $\sim 70$~days of
the X-ray deep minimum.

In the present paper, we wish to study the He~II~$\lambda 4686$\AA\ emission line
detected by Steiner \& Damineli (2004, hereafter SD04) and by Martin et al.\ (2006, hereafter M2006).
In particular, we try to explain its sharp rise and subsequent decline around the \etc\ periastron.
M2006 attribute this to a temporary mass-ejection or wind-disturbance on the primary star.
We instead, attribute the emission to the wind of the secondary star, and the variation
in the He~II~$\lambda 4686$\AA\ emission to variation in the properties of the
secondary wind, and to a collimated fast wind (CFW) blown by the secondary star.
The onset of accretion will account in our model for the disappearance
of the He~II~$\lambda 4686$ line during the X-ray minimum, when the
secondary wind is shut down.
In Section 2, we examine the possible influence of the X-rays emitted by the
shocked winds on the secondary wind.
The general properties of the He~II~$\lambda 4686$ line
and the outline of our proposed explanation are described in Section 3.
Our summary and predictions are presented in Section 4.

% ==========================================================
\section{X-RAY ILLUMINATION OF THE SECONDARY WIND}
\label{sec:xray}
% ==========================================================

% ================================
\subsection{The Binary system}
\begin{figure}
%%%\resizebox{0.7\textwidth}{!}{\includegraphics{etahef1.eps}}
\resizebox{0.7\textwidth}{!}{\includegraphics{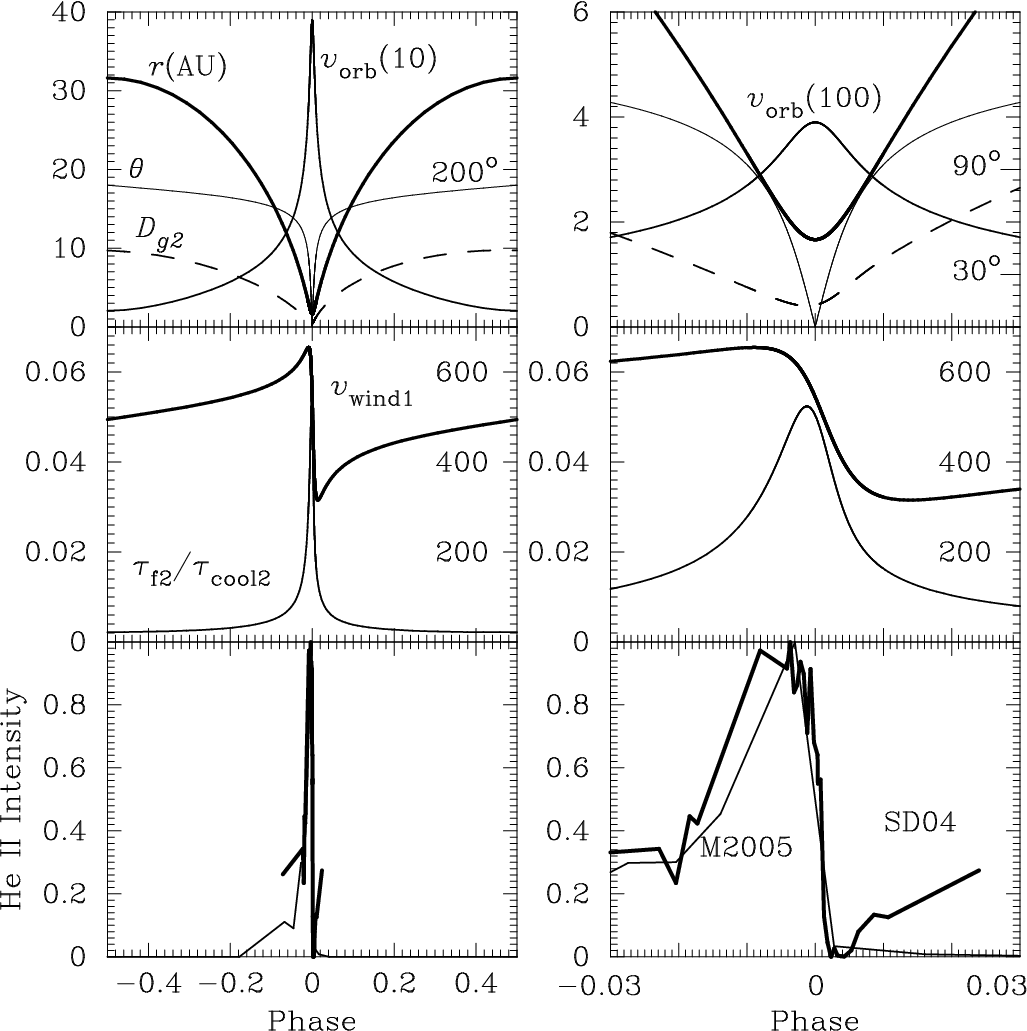}}
%%\epsfysize=8cm
%%\epsffile{etaheb1.eps}
%{\includegraphics{etahef2.eps}}
\caption{
Several physical variables as function of orbital phase.
The left column covers the entire orbit, while the column on the right
covers the time just prior and after periastron.
In the upper two rows quantities are calculated from the model and
phase zero is exactly at periastron.
Upper row: The orbital separation (in AU), the distance of the stagnation
point from the secondary $D_{g2}$ (dashed line; AU), the orbital angle $\theta$,
and the relative orbital speed of the two stars (in $10 \km \s^{-1}$
on the left and $100 \km \s^{-1}$ on the right).
The angle $\theta$ is the relative direction of the two stars as measured
from periastron (scale on the right in degrees).
Second row: Thick line: The velocity of the primary wind relative to the secondary
$v_{\rm wind1}$; scale on the right side (km~$\s^{-1}$).
Thin line: The ratio of the flow time of the shocked secondary wind
out of the colliding wind region $\tau_{f2}$ to its radiative cooling time
$\tau_{\rm col2}$.
Lower row: The intensity of the He II $\lambda~4686$ line.
The thick line depicts the equivalence width as given by SD04 for
the time period when measurements are reliable, while the thin line is the
He~II~$\lambda 4686$ line intensity from M2006.
Both lines are scaled to their maximum value.
In the lower row phase zero is at June 29, 2003.
}
\label{binf}
\end{figure}

% ================================
The \etc\ binary parameters are as in the previous two papers in this series
(Soker 2005b; Akashi et al.\ 2006).
The assumed stellar masses are $M_1=120 M_\odot$, $M_2=30 M_\odot$, the eccentricity is
$e=0.9$, and orbital period 2024 days, hence the semi-major axis is
$a=16.64 \AU$, and periastron occurs at $r=1.66 \AU$.
The mass loss rates are $\dot M_1=3 \times 10^{-4} M_\odot \yr^{-1}$
and $\dot M_2 =10^{-5} M_\odot \yr^{-1}$.
The primary's wind profile is $v_1=500 [1-(0.4 \AU/r_1)] \km \s^{-1}$,
where $r_1$ is the distance from the center of the primary.
In the colliding wind region relevant to us, we can set $r_1=r$, where $r$
is the orbital separation between the two stars.
The secondary's terminal wind speed is taken to be $v_2=3000 \km \s^{-1}$.
The orbital separation $r$, the relative orbital velocity of the two stars
$v_{\rm orb}$, the relative angle of the two stars measured from periastron,
and the distance of the stagnation point of the colliding two winds from the
secondary $D_{g2}$, are plotted on the first row of Figure \ref{binf}.
In the second row the velocity of the primary wind relative to the stagnation
point $v_{\rm wind1}$ is depicted by the thick line.
The thin line represents the ratio of $\tau_{f2}/\tau_{\rm cool2}$,
where $\tau_{f2}$ is the typical time for the shocked secondary wind to flow
out of the shocked region (winds interaction zone), while $v_{\rm cool2}$ is the
radiative cooling time of the shocked secondary wind.
For more detail on these quantities see Soker (2005b) and Akashi et al.\ (2006).

The secondary can be assumed to be an O star.
Somewhat evolved main sequence O-stars with $M_2=30 M_\odot$
can have an effective temperature of $T_{\rm eff} \simeq 40,000 \K$, and
a luminosity of $L_2 \simeq 3 \times 10^5 L_\odot$, hence a
radius of $R_2 \simeq 11 R_\odot$; such stars have mass loss rates of up
to $\sim 10^{-5} M_\odot \yr^{-1}$ (e.g., Repolust et al. 2004).
These estimates are associated with large uncertainties
since most likely the secondary underwent a massive accretion event
$\sim 160$~yr ago (Soker 2001) and hence it is likely to be out of thermal
equilibrium.
Recently, Verner et al. (2005) deduced the following secondary properties:
$T_{\rm eff} \simeq 37,200 \K$, $L_2 \simeq 9.3 \times 10^5 L_\odot$,
$R_2 \simeq 23.6 R_\odot$, $v_2=2000 \km \s^{-1}$, and
$\dot M_2 \simeq 8.5 \times 10^{-6} M_\odot \yr^{-1}$.
We will therefore scale the model with a  value of $R_2 = 20 R_\odot$.

In the first two papers in the series (Soker 2005b; Akashi et al. 2006)
the acceleration zone of the secondary's wind was not considered, and it
was assumed that the secondary wind encounters the shock wave at
its terminal velocity.
In fact, because of the non-negligible acceleration zone, the secondary wind
does not reach its terminal speed when encounter the shock wave.
This in turn reduces the distance of the stagnation point to the secondary, $D_{g2}$,
and enhances the effect of accretion of the shocked primary wind near the stagnation
point by the secondary.
In the present paper, we treat the acceleration zone more carefully,
as we believe it is in this region that the He~II~$\lambda 4686$ line is formed,
{{{ a possibility raised already by SD04. }}}
We adopt the $\beta$ velocity profile for the secondary's wind acceleration zone:

\begin{equation}
v_{2r}(r_2)= v_2\left(1-\frac{R_2}{r_2}\right)^\beta
\label{vbeta}
\end{equation}
with $\beta=1$, where $r_2$ is the distance from the center of the secondary,
and $v_2=3000 \km \s^{-1}$.
For the parameters assumed here, the gravitational acceleration on
the surface of the secondary is $\log [g_2({\rm cm} \s^{-2})]=3.3$.
In such stars, Venero at al. (2002)
%, whom later we cite their results
%on the emission of the He~II~$\lambda 4687$ line,
find more efficient acceleration, hence a narrower acceleration zone,
which implies a lower column density through the wind with less
X-ray absorption.
The density profile is given by mass conservation, and for the parameters
used here, $\dot M_2=10^{-5} M_\odot \yr^{-1}$, it is
\begin{equation}
\rho_{2} (r_2)= 9 \times 10^{-14}   % 8.63
\left( \frac{r_2}{R_2} \right)^{-2}
\left( \frac{R_2}{20 R_\odot} \right)^{-2}
\left(1-\frac{R_2}{r_2}\right)^{-1} \g \cm^{-3}.
\label{rho2}
\end{equation}

The column density from infinity inwards to the secondary wind at radius $r_2$
is given by integrating over the proton number density $n_{p2}=\rho_{2}/m_p$ \,
\begin{equation}
N_{p2}=\int^{\infty}_{r_2} n_{p2} dr_2^\prime =
- 5 \times 10^{22}
\left( \frac{R_2}{20 R_\odot} \right)^{-1}
\ln \left( 1-\frac{R_2}{r_2}\right) \cm^{-2}.
\label{N2}
\end{equation}
As mentioned above, the acceleration of the secondary wind is likely to be larger
than that given by equation (\ref{vbeta}), hence the density
will be lower than that given by equation (\ref{rho2}), and the column
density lower than that of equation (\ref{N2}).

\begin{figure}
\resizebox{0.66\textwidth}{!}{\includegraphics{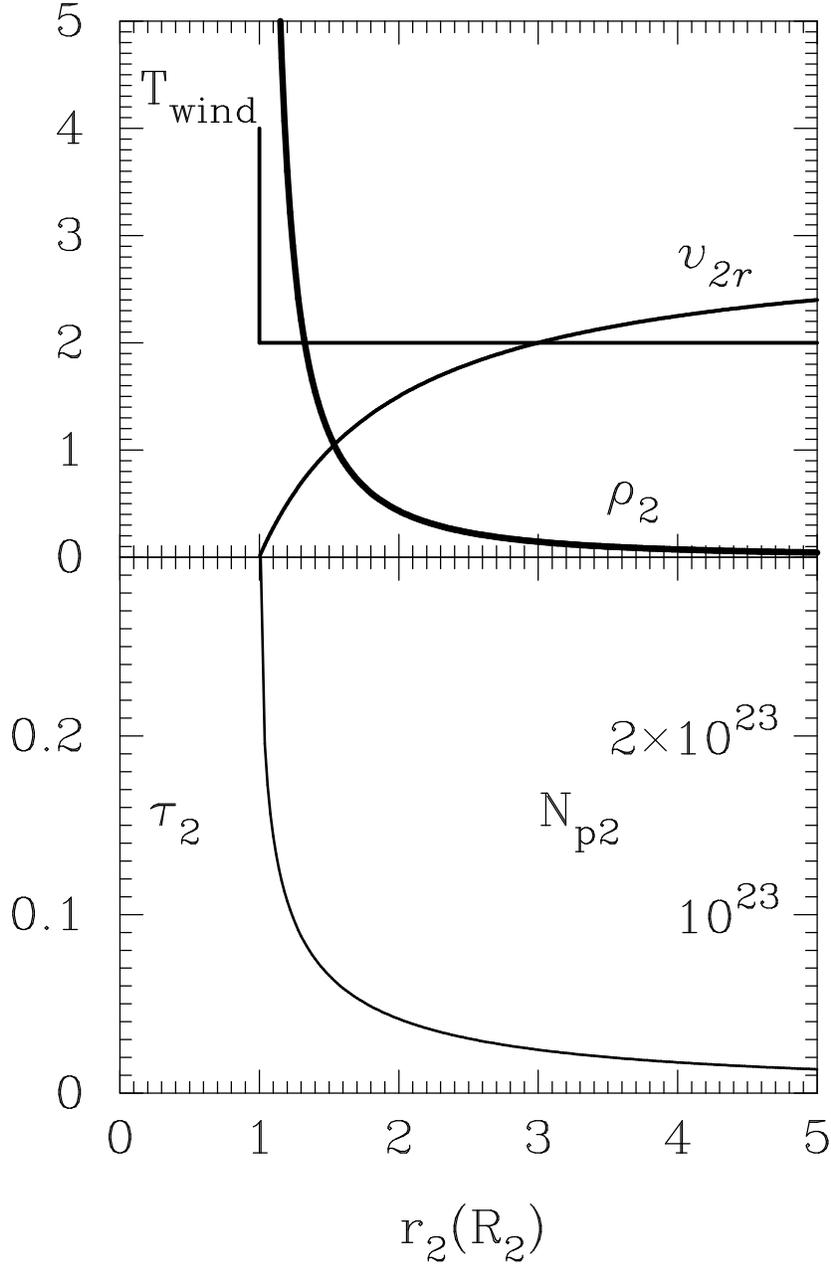}}
\caption{ Upper panel: The assumed secondary wind parameters: wind
velocity $v_{2r}$ (in units of $1000 \km \s^{-1}$), temperature
$T_{\rm wind}$ (in units of $10000 \K$), and the density
$\rho_{2}$ (in units of $10^{-13} \g \cm^{-1}$) for a mass loss
rate of $\dot M_2=10^{-5} M_\odot \yr^{-1}$. The distance from the
secondary center is in units of secondary radius, taken to be
$R_2=20 R_\odot$. Lower panel: The optical depth for radiation
coming from infinity $\tau_2$ (units on the left side), and the
proton column density $N_{p2}$ (in units of $\cm^{-2}$ on the
right side). Note that the same line represents both $\tau_2$ and
$N_{p2}$. } \label{accf}
\end{figure}

The He~II~$\lambda 4686$ line intensity very sensitively increases with increasing wind's
temperature $T_0$.
In the different relevant models of Venero et al. (2002),
the wind temperature $T_{\rm wind}=T_0$ spans the range $15,000-32,000 \K$.
Consequently, we assume a fixed temperature of $T_0 = 20,000 K$.
The velocity, density, and the temperature profiles of the wind model are plotted in
the upper panel of Figure \ref{accf}. The bottom panel shows the column density
and optical depth into the wind.

% ================================
\subsection{X-ray Emission}
% ================================
The assumptions entering the calculation of the X-ray emission
were summarized by Akashi et al. (2006), who were interested in the
X-ray emission in the $2-10 \kev$ band. Here too, we need to know
the X-ray flux from both the primary and secondary shocked winds.

% =================================
\subsubsection{Primary wind}
\begin{figure}
\resizebox{0.66\textwidth}{!}{\includegraphics{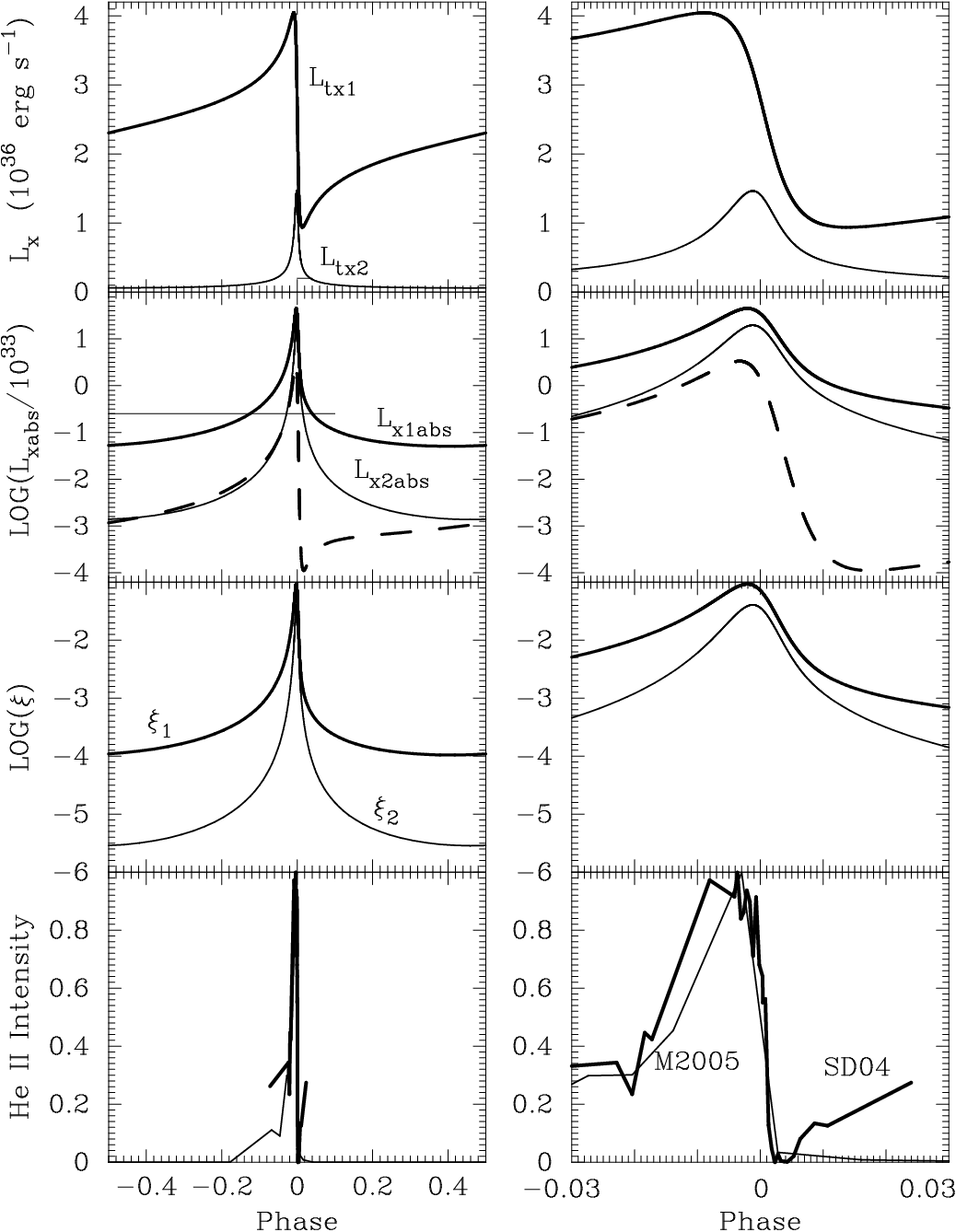}}
%{\includegraphics{eta46f3.eps}}
\caption{
Upper row: The total X-ray luminosity of the post-shock winds,
the primary wind $L_{tx1}$ (thick line), and the secondary wind
$L_{tx2}$  (thin line), both in units of $10^{36} \erg \s^{-1}$.
The time that the secondary wind does not exist in the accretion model
for the minimum is marked by a short horizontal line in the left panel
(in the orbital-phase range $0-0.04$).
Second row: The fraction of the total X-ray emission
that would be absorbed by the secondary star, for $R_2=20 R_\odot$, if
no absorption in the secondary wind would occur.
The dashed line shows the contribution of the X-ray emitted
by the primary post-shock wind above $1 \kev$. All three lines
are in units of $10^{33} \erg \s^{-1}$ and in logarithmic scale.
The long horizontal line is drawn to emphasize the stronger X-ray emission
before periastron compared with post periastron.
Third row: The ionization parameter defined in equation (\ref{xi})
calculated close to the secondary photosphere at $r_2=1.05 R_2$,
where the nucleon number density is  $n_n \simeq 10^{12} \cm^{-3}$,
and for $r_x=D_{g2}$; in logarithmic scale and units of $\erg \cm \s^{-1}$.
Lower row: The He II $\lambda~4686$ line intensity as in lower row of Figure 1.
}
\label{lxf}
\end{figure}

% =================================
We are interested in the X-ray emission close to minimum, i.e., periastron passage,
where the shocked primary's wind is very dense and its cooling
time is much shorter than the flow time (e.g., Pittard \& Corcoran 2002; Soker 2003).
Therefore, we can assume that the post-shock primary's wind material near the
stagnation point cools instantaneously by emitting all its thermal energy.
During the pre X-ray minimum the primary wind speed relative to the
stagnation point is $v_{\rm wind1} \simeq 600 \km \s^{-1}$, corresponding
to a post-shock temperature of $T_{s1} \simeq 5 \times 10^6 \K$, while after the
X-ray minimum the pre-shock primary wind speed is $v_{\rm wind1} \simeq 400 \km \s^{-1}$
corresponding to post-shock temperatures of $T_{s1} \simeq 2.2 \times 10^6$ K
(this is evident by the dashed line in figure \ref{lxf} to be discussed later).

$~$From the shape of the shock front of the primary wind, Akashi et al.\ (2006)
estimate the fraction of the primary wind that is strongly shocked
and contributes most to the X-ray emission to be $k_1 \simeq 0.1$.
The total X-ray emission of the shocked primary wind is therefore
\begin{equation}
L_{tx1} \simeq  3.4 \times 10^{36}
\left( \frac {k_1}{0.1}\right)
\left( \frac{\dot M_1}{3 \times 10^{-4} M_\odot \yr^{-1}}\right)
\left( \frac{v_{\rm wind1}}{600 \km \s^{-1}}\right)^2 \erg \s^{-1}.
\label{lx1}
\end{equation}
The value of $L_{tx1}$ as a function of orbital phase is drawn by the thick
line in the first row of Figure \ref{lxf}.

Assuming that the emission is concentrated near the stagnation point
and neglecting absorption by the material near the stagnation point,
we derive the radiation power that is absorbed by the secondary
$L_{\rm x1abs}=0.5L_{tx1}(R_2/D_{g2})^2$.
The variation of $L_{\rm x1abs}$ with orbital phase for $R_2=20 R_\odot$
is drawn by the thick line in the second row of Figure \ref{lxf}.
The dashed line in that plot shows the contribution of X-rays above $1~$kev
to $L_{\rm x1abs}$.
In order to make the asymmetry around periastron passage clear, we draw an
horizontal line at the absorbed power value of $2.5 \times 10^{32} \erg \s^{-1}$.
This shows that the X-ray emission is stronger before the spectroscopic event
than after it.

% =================================
\subsubsection{Secondary wind}
% =================================
The radiative cooling time of the post shock secondary wind material
$\tau_{\rm cool2}$ is much longer than its flow time out of the
wind-collision region.
Like Akashi et al. (2006), we assume that about half of the mass blown
by the secondary star is shocked in a perpendicular shock front,
and that the total emitted energy is a fraction
$k_2 \tau_{f2}/\tau_{\rm cool2}$ of the thermal energy of the post
shock gas.
Here $\tau_{f2} \equiv D_{g2}/v_2$ is the characteristic flow time
of the shocked wind out of the interaction region (Soker 2005b).
The value of $\tau_{f2}/\tau_{\rm cool2}$ is plotted
in the second row of Figure 1.
By fitting the X-ray luminosity to observations,
Akashi et al. (2006) obtain $k_2 = 2$.
The contribution of the shocked secondary wind material to the
X-ray luminosity is therefore
\begin{equation}
L_{tx2} = 1.4 \times 10^{35}
\left( \frac{\dot M_2}{10^{-5} M_\odot \yr^{-1}} \right)
\left( \frac{v_2}{3000 \km \s^{-1}}\right)^2          % 6.526 \times 10^{36}
\left( \frac {k_2 \tau_{f2}/\tau_{\rm cool2}}{0.01}\right)  \erg \s^{-1}.
\label{lx22}
\end{equation}
This is much smaller that $L_{tx1}$. However, the secondary wind velocity and thus
the post-shock temperatures are much higher, corresponding to $T_{s2}=1.3 \times 10^8 \K$.
Consequently, the X-ray radiation produced by the secondary wind
is much harder and its transmission through the primary wind much higher.
The value of $L_{tx2}$ as a function of orbital phase with $k_2=2$ is drawn
by the thin line in the first row of Figure \ref{lxf}.

Assuming that the emission is concentrated in the region
approximately $D_{g2}$ away from the secondary, and neglecting
absorption by the shocked secondary wind, we derive the radiation
power of the secondary wind that is absorbed by the secondary star
$L_{\rm x2abs}=0.5L_{tx2}(R_2/D_{g2})^2$. The variation of $L_{\rm
x2abs}$ with orbital phase for $R_2=20 R_\odot$ is drawn by the
thin line in the second row of figure \ref{lxf}. {{{ SD04
already noted that the sharp rise in the X-rays and its absorption
by the secondary star might be the reason for the rapid rise in
the He~II $\lambda$4686 line intensity. }}}

% ============================================
\subsection{Influence of X-rays on Secondary Wind}
% ============================================
% ============================================
\subsubsection{Wind velocity}
% ============================================
The influence of X-rays incident upon an O-star wind was studied by
Stevens \& Kallman (1990; hereafter SK90) in the context of high mass X-ray binaries.
For the typical physical values appropriate for \etc, the results of SK90 imply that the
X-rays do not penetrate the wind deep enough to influence the mass loss rate.
However, the X-rays do ionize the wind, and by that
reduce the efficiency of the radiative acceleration of the wind.
Hence, with the presence of X-rays, the wind in the acceleration zone is slower and denser.
The important parameter that determines the wind ionization by X-rays is the
photoionization parameter, which is define as (SK90)
\begin{equation}
\xi=\frac{L_x}{n_n r_x^2} = 4.5 \times 10^{-3}    % 4.468
\left( \frac {L_x}{10^{36} \erg \s^{-1}} \right)
\left( \frac {n_n}{10^{12} \cm^{-3}} \right)^{-1}
\left( \frac {r_x}{1 \AU} \right)^{-2} \erg \cm \s^{-1},
\label{xi}
\end{equation}
where $r_x$ is the distance from the X-ray source and $n_n$ the
nucleon number density. We are interested in the effect of
photoionization by the primary and secondary shocked winds on gas
just off the surface of the secondary, where its wind is launched.
Thus, we take $r_x=D_{g2}$ and $n_n= 10^{12}$ cm$^{-3}$
(appropriate for $r_2=1.05R_2$, c.f., equation (\ref{rho2})).
%$n_n=8\times 10^{11}$ cm$^{-3}$
The resulting ionization parameter as a function of orbital phase
is drawn in the third row of Figure \ref{lxf}
for photoionization due to both the primary ($\xi _1$, thick line)
and secondary ($\xi _2$, thin line) shocked winds.

Learning from the results of SK90 (their figure 8 and table 2) who use a
temperature of 10~keV for the X-ray emitting gas, which is similar to that of
the shocked secondary wind, we find that for $r_2 \sim 1.5 R_2$, a typical region
where the He~II~$\lambda 4686$\AA\ emission forms,
the X-rays start slowing down the secondary wind when $\xi \simeq 10^{-4}$ erg~s$^{-1}$~cm.
For $\xi=10^{-3}$ erg~s$^{-1}$~cm, the wind speed is $\sim 10\%$ lower than
its undisturbed velocity, while for $\xi=3 \times 10^{-3}$ erg~s$^{-1}$~cm,
it is $\sim 25 \%$ slower.
All the values  of $\xi$ cited above are calculated at $r_2=1.05R_2$,
(the relevant emitting region is at $r_2 \sim 1.5 R_2$), as required for
comparing with the results of SK90.
The mass loss rate does not change.
Therefore, without considering other effects, the third row in
Figure \ref{lxf} suggests that the secondary wind will slow down
and become denser approximately between phase -0.05 and +0.04.
We postulate that at this time, the He~II~$\lambda 4686$ line intensity
will be enhanced.

When X-ray absorption by the secondary wind is taken into account, the effect of the
soft X-rays from the shocked primary wind (i.e., $\xi _1$) is eliminated.
The proton column density from infinity to the relevant region $r_2 \simeq 1.5 R_2$
is $N_{p2} \sim 5 \times 10^{22} \cm^{-2}$.
For this column density, practically all X-ray
emission below $\sim 1 \kev$ (e.g., primary shocked wind)
will thus be absorbed before reaching this region.
Only the much harder X-rays emanating from the shocked {\it secondary} wind
can reach $r_2 \simeq 1.5 R_2$ and effect the secondary wind by means of $\xi _2$.
SK90 consider both absorption and harder X-ray radiation.
To estimate the influence of absorption on the secondary wind X-rays
we compare Figure 6 of SK90 with their Figure 1,
and for harder X-ray emission their Figure 7 with Figure 1.
Crudely, a column density of $N_{p2} \sim 5 \times 10^{22} \cm^{-2}$
reduces $\xi _2$ by a factor of $\sim 20$.
Our assumptions, on the other hand underestimate $\xi$:
($i$)
The optical depth should actually be calculated from $r_2=D_{g2}$ inward, making its real
value smaller, and the influence of the X-ray emission higher.
However, for our goal it is adequate to use the column density from infinity and to avoid
unnecessary complications, remembering that we overestimate X-ray absorption by
the secondary wind.
($ii$) We also underestimate the influence of the X-ray emission by taking the X-ray
source to be at the stagnation point, at a distance $D_{g2}$ from the center
of the secondary.
The secondary wind shock wave is actually closer to the secondary than the
contact discontinuity surface, by a factor of $\sim 2$ (Akashi et al.\ 2006).
Over all, we estimate that the third row of Figure \ref{lxf} overestimates
$\xi_2$ by a factor of $\sim 3-10$ ($\xi_1$ is negligible after absorption).
Considering the many uncertainties involved, we cautiously propose that
as periastron passage approaches, say after orbital phase $\sim -0.05$, i.e.
100 days within periastron passage, the X-ray emission from the shocked wind
can slow down the secondary wind by up to a few 10\%.

% ============================================
\subsubsection{Wind temperature}
% ============================================

Venero et al. (2002) find that the He~II~$\lambda 4686$ emission line intensity
is sensitive to the wind temperature and increases as the wind temperature
increases and/or as the region of maximum temperature in the wind moves outward.
We therefore examine the heating of the wind by the X-ray.
We neglect heating by the optical+UV energy flux of the primary star because
the primary optical+UV spectrum is similar to that of the secondary star, but
its flux is much smaller than the secondary flux in the relevant region at
$r_2 \sim 1.5 R_2$.
The X-ray emission, on the other hand, is much harder and it
ionizes the gas and heats it.

As can be seen in Figure (15) of Venero et al. (2002) an increase by
$\sim 1000 \K$ of the wind temperature can substantially increase
the He~II~$\lambda 4686$ emission line intensity.
Say a fraction $\delta$ of the X-rays absorbed by the secondary wind goes to heat.
The particle number loss rate by the wind in the hemisphere
facing the X-ray source is $\dot M_2/(2\mu m_H)$, where
$\mu m_H$ is the mean mass per particle.
Therefore, the X-rays raise the wind temperature by
\begin{equation}
\Delta T_w \sim \delta \frac {2}{3k_B} \frac {L_{\rm xabs}}{\dot M_2/(2\mu m_H)} =
1500 ~ \frac {\delta}{0.1}
\left( \frac {L_{\rm xabs}}{10^{33} \erg \s^{-1}} \right)
\left( \frac {\dot M_2}{10^{-5} M_\odot \yr^{-1}} \right)^{-1} \K,
\label{tw}
\end{equation}
where $k_B$ is the Boltzmann constant.
Near the He~II $\lambda$4686 maximum ${L_{\rm xabs}} \sim {10^{33} \erg \s^{-1}}$, and
even for $5 \%$ efficiency, $\delta \sim 0.05$, the heating of the
wind might be important.
However, it seems that raising the wind temperature cannot be the main effect
causing the variation in the He~II $\lambda$4686 line intensity.

% ==========================================================
\section{THE PROPOSED MODEL FOR THE He II $\lambda 4687$\AA\ LINE}
\label{sec:heii}
% ==========================================================
\subsection{The He II $\lambda 4686$\AA\ Line in $\eta$ Carinae}
%=====================================================================

The general binary model for our purposes is described in the first two
papers of the series (Soker 2005b; Akashi et al. 2006).
The dependence of some relevant binary parameters on the orbital phase,
with phase zero at periastron, are drawn in the first two rows of Figure 1.
The third row of Figure 1 shows the behavior of the He~II~$\lambda~4687$
emission line.
The thin line depicts the He II line intensity as given by M2005,
while the thick line is the equivalent width from SD04, giving only
the time period when measurements are reliable (Damineli, A. 2005,
private communication).
Both lines are normalized to their maximum intensity.
Following the discussion by M2005 of the problematic continuum assessment of SD04,
we will refer more to the He~II~$\lambda 4686$ intensity as given by M2005,
where there is no rise in intensity after the peak.
Note that in our modelling, phase zero is well defined at periastron.
This is not to be confused with phase zero defined from observations of
the intensities of different lines (June 29, 2003; JD=2,452,819.8); in the
latter case phase zero is assumed to be near periastron, but it is not
well defined (see a footnote in M2005).

Three qualitative regimes are seen in the intensity evolution.
($i$) {\it No emission.} During most of the orbit the He~II
intensity is very weak, practically zero, {{{ although we note that
M2006 discuss the slight possibility that Thackeray (1953) detected the
line in weak emission. }}} ($ii$) {\it Slow rise.} At orbital phase $\sim
(-0.1)$ ($\sim 200$~days before periastron passage) the intensity
starts to rise slowly. M2005's first detection is $\sim 140$~days
before the spectroscopic event (in the lower row of Figure 1 we
connected the observation points of M2005, and the previous
measurement to the first detection 220 days earlier). {{{ The
staring point of the slow rise is the main discrepancy between
SD04 (slow rise started almost a year before minimum) and M06
(slow rise started $\sim 5$ moths before minimum). }}} ($iii$) {\it
Peak.} At orbital phase $\sim -0.02$ ($\sim 40$ days before
periastron passage) the intensity sharply rises to its maximum
value at phase $\sim -0.006$ after which it
sharply declines back to its minimum value {{{ (SD04).
With inferior temporal resolution, M2006 quote a phase of $-0.004$ for the peak.}}}
The He~II peak is not as well located as the narrow X-ray flares, but it seems to occur
when the X-ray intensity is dropping (M2005). Namely, as the X-ray
source is shut-down, the He~II~$\lambda 4686$ line gains its
intensity. Then both X-ray and He~II~$\lambda 4686$ intensity are
at their minimum.

After correcting for extinction by a factor of about 100,
SD04 find the peak luminosity of the He~II~$\lambda~4686$\AA\
line to be $L_{\rm HeII} \sim 100 L_\odot$, which amounts to
$\dot N _{\rm HeII} \sim 9 \times 10^{46}$ photons s$^{-1}$
in the line alone.
M2006 find the peak luminosity to be $\sim 2.5$ times higher at
$1.4 \times 10^{36} \erg \s^{-1}$.
{{{ Since M2006 did not observe \etc\ at the exact maximum, the true
He~II~$\lambda~4686$\AA\ maximum might be even higher than this. }}}
Noting that the X-rays and He~II~$\lambda 4686$ line rise together before the X-ray
luminosity drops, SD04 suggest that the X-rays can account directly for the
He II emission by ionizing the helium.
SD04 extrapolate the X-ray spectrum in the energy range $1 \kev < E_\gamma < 10 \kev$
as given by Corcoran and collaborators (Ishibashi et al. 1999; Corcoran et al. 2001;
Pittard \& Corcoran 2002) down to $E_1=54 \ev$ the ionization threshold of He$^+$.
{{{ We used the APEC plasma compilation (Smith et al. 2001) to calculate the
photon flux from 54~eV to 10~keV
% (e.g., Pittard et al. 2005),
for plasma at the temperature of $\sim 5 \times 10^6 \K$ appropriate for the shocked
primary wind (the secondary wind supplies a much smaller number of photons).
We get $1.7\times 10^9$~photons~$\erg^{-1}$ in this range
(as would be expected $\sim$ photon/keV).
At maximum, the X-ray luminosity of the shocked primary wind is
$L_{tx1} \simeq 4 \times 10^{36} \erg \s^{-1}$, which implies an ionizing photon
number rate of $\sim 7 \times 10^{45} \s^{-1}$.
Before even considering the solid angle occupied by the secondary wind and the
efficiency of He~II~$\lambda~4686$\AA\ line formation,
there is an order of magnitude deficiency in photons. }}}
When considering the efficiency of line formation the number of ionizing
photons from the X-ray emitting gas is found to be more than
two orders of magnitude below the required ionization flux
to explain the He II $\lambda~4686$\AA\ line.
Therefore, as already noted by M2006, another explanation is
required for the He~II~$\lambda 4686$ emission.

Another problem with the X-rays directly ionizing the He is that the
observed He~II~$\lambda 4686$ line maximum occurs 18 days after the {{{ observed
X-ray flux starts to decline to its deep minimum.
This point was raised by M2006 and we reinforce it here by arguing that this decline
in X-ray flux must be intrinsic to the X-ray source.
Indeed, in Akashi et al. (2006), where we modeled
the X-ray light curve of \etc, we found that the decrease in X-ray luminosity
at phase $-0.015$ ($\sim 30$~days before periastron) has to be intrinsic to the X-ray
source and can not be explained by absorption towards our line of sight.
The results of Akashi et al. (2006), thus, rule out the possibility that the He gas
sees a strong steady X-ray source while we observe the deep X-ray minimum.
In summary, X-ray ionization of the He gas appears to be highly unlikely.}}}

{{{ Both SD04 and M2006 find that the He~II~$\lambda~4686$\AA\ line
is blue shifted. SD04 have better time sampling and find the line of sight
velocity to rise from $\sim 100 \km \s^{-1}$ to $\sim 400 \km \s^{-1}$ at its
maximum intensity. }}}
Stahl et al.\ (2005) observed the reflection of emission from
the polar direction and consider the He~II~$\lambda~4686$\AA\ line to
be formed in a shock front.
They too find the He~II~$\lambda~4686$\AA\ line to be blue shifted.
This behavior of the He~II~$\lambda~4686$\AA\ line and its rapidly evolving blue-shift
show that this velocity cannot be attributed to the orbital motion of the secondary star,
and it must originate from genuinely outflowing gas perhaps in a biconical flow.
{{{ The redshifted part of the flow then must be obscured.
In our model, the He~II~$\lambda~4686$\AA\ line is formed close to the secondary star.
Therefore, the star itself will block any red-shifted emission if it exists. }}}

M2006 discuss three possibilities for the He II line emission:
In the colliding wind region; in a dense shell ejected by the primary star
during the early spectroscopic event; or a combination of the two.
In any case, it seems an enhanced mass loss rate by the primary is required
in their suggestion.
M2006 require the enhanced mass loss rate from the primary
to temporarily be several times $10^{-3} M_\odot \yr^{-1}$ for a wind speed of
$v_1=700 \km \s^{-1}$.
They note that in their suggested scenario for the He~II~$\lambda~4686$\AA\
line the energy supply appears to be marginal, and requires radiative
processes to enhance the He~II~$\lambda 4686$ line formation.
In summary, we find neither the explanation by SD04 nor those by M2006
for the He~II~$\lambda~4686$\AA\ emission to be realistically satisfactory.

% ==========================================================
\subsection{The Proposed Model}
%=====================================================================

Many O stars are known to have strong emission in the
He~II~$\lambda 4686$ line, with an equivalent width of $1-3$~\AA
(e.g. Grady et al.\ 1983), which translates into a line luminosity
of $\sim 10 L_\odot$.
The stars in the sample of Grady et al.\ (1983) have mass loss rates
lower by at least a factor of 5 compared to the secondary in the \etc\
system, and their bolometric luminosity is $\simeq L_2$
as given be Verner et al. (2005).
The He~II~$\lambda 4686$ line in the sample of Grady et al.\ (1983)
displays velocities in the range of $0-600 \km \s^{-1}$.
Similar blue-shifted velocities were inferred by SD04 for \etc.
Grady et al.\ (1983) argue that the He~II~$\lambda 4686$ line intensity variability
is connected to changes in the entire wind acceleration zone.
The changes in the acceleration zone can be in the ionization balance,
and the mass loss rate (e.g., Venero et al.\ 2002).
Kunasz (1980) found that when the He~II~$\lambda 4686$ line is only slightly
in emission it is very sensitive to the wind parameters.
This is most likely the case in $\eta$ Car, where the He II $\lambda 4686$
does not exist most of the time, and appears only several months before
the spectroscopic event (M2006).
Late WNL stars, whose hydrogen abundance is low by a factor
of a few relative to solar, are also known to be strong He~II~$\lambda 4686$
emitters, with line luminosity of up to several$\times 100 L_\odot$ (Crowther 2000).
{{{ This sensitivity of the He~II~$\lambda 4686$ to wind parameters under
particular conditions, implies that the variations in this line might come
with unnoticeable variations in other lines.
Furthermore, time variability in some lines in WR stars are known to be unrelated
to variability in the He~II~$\lambda 4686$ line, e.g., NV~$\lambda 4945$ and
C~IV~$\lambda 5806$ (Morel et al. 1999).
In a sample of seven O stars Ninkov et al. (1987) find the variation
in the equivalence width of the C~III~$\lambda 5696$ line from star
to star to be unrelated to the variation in the equivalence width of the
He~II~$\lambda 4686$ line.
The variation of the N~III~$\lambda \lambda 4634-41$ (multiplet)
is also weakly related to the He~II~$\lambda 4686$ line.
Different lines are formed at different locations in the wind,
and therefore are influenced differently as the system approaches periastron.
Indeed, Rauw et al. (2001) studied the O7.5I+ON9.7I binary system HD~149404, and found
the He~II~$\lambda 4686$, C~III~$\lambda 5696$, N~III~$\lambda \lambda 4634-41$,
N~III~$\lambda \lambda 5932,5942$ and S~IV~$\lambda \lambda 4486,4504$
lines to originate from different locations in the binary system.
Rauw et al. (2005) find temporal and spacial correlations between the He~II~$\lambda 4686$
line and H$\alpha$ in the very massive binary system WR20a (WN6ha + WN6ha).
We therefore do not necessarily expect variability in the intensity of all other lines
to follow that of the He~II~$\lambda 4686$ line, although it is possible that some
lines show qualitatively similar behavior. }}}

Comparing the properties of these O stars from the literature and that of
$\eta$ Car, we suggest that the He~II~$\lambda 4686$ line is formed
in the acceleration zone of the secondary wind.
{{{ That the He~II~$\lambda~4686$\AA\ line originates in the acceleration zone
of the secondary wind has been proposed by SD04.
However, there is a fundamental difference between our model and their's.
In our model, the energy source for the He line is
the outflowing secondary wind and not the X-ray emission as suggested by SD04. }}}
The energy in the He~II line is negligible compared with the kinetic
energy of the wind, and as with the O stars mentioned above, the
energy in the He~II~$\lambda 4686$ line results from excitation and
ionization processes within the accelerated wind.
Therefore, there is no energy budget crisis in our model.
The assumption that the He~II line results in the acceleration zone of
the secondary wind accounts for the behavior of the He~II~$\lambda 4686$
emission line as follows.

($i$) {\it No emission.} During most of the orbit the conditions in the
secondary wind are such that the line is very weak. {{{ (Theoretical
calculations by Kunasz (1980) show that the line might even appear }}}in absorption.)
However, the conditions are such that a slight increase in the density
and/or temperature in the acceleration zone will lead
He~II~$\lambda 4686$ to show up in emission.

($ii$) {\it Slow rise.}
As the two stars approach each other the ionization parameter ($\xi _2$) due
to the secondary X-rays increases such that enough X-ray penetrates
to the acceleration zone, i.e., the ionization parameter in the acceleration
zone of the secondary wind exceeds 10$^{-4}$ erg~s$^{-1}$~cm.
Consequently, the wind speed decreases while the mass loss rate is unaffected,
as discussed in section 2.3.1.
This results in a higher density in the acceleration zone, which based on the
results of Kunasz (1980), increases the intensity of the He~II~$\lambda 4686$
emission line.
In addition, the X-rays raise the wind temperature by several$\times 100 \K$
(section 2.3.2).
This slightly higher wind temperature also increases the He~II~$\lambda 4686$
line intensity (Venero et al. 2002).
We note that it is sufficient for the mass loss rate (Kunasz 1980) or
the temperature in the wind's acceleration zone (Venero et al. 2002) to
increase slightly in order to substantially strengthen the He~II line.
The full modelling of these processes require a stellar structure code and are
much beyond the scope of the present paper,
but the results of the previous section do suggest that the X-ray emission from
the colliding wind can influence the wind speed and temperature, which in turn
can drastically change the He~II~$\lambda 4686$ line intensity.

Another effect, not treated in the previous section, is the possibility
that dense blobs are accreted from the post-shock primary wind region (Soker 2005b).
As these blobs fall into the secondary wind acceleration region, they will form
shock waves in the secondary wind. This will further heat the wind and in a limited
region would further enhance the He~II line.

($iii$) {\it Peak.} According to our model, for $\sim 70$~days,
the secondary accretes from the primary wind (Soker 2005b) after the
collapse of the shocked primary wind material in the region of the
stagnation point onto the secondary.
As we show below, this gas has high specific angular momentum.
The collapse starts at phase $\sim -0.02$, namely, $\sim 40$ days before periastron
passage (Akashi et al. 2006); the X-ray minimum starts at phase $\sim 0$.
Since most of the gas is first accreted near the equatorial plane around the secondary,
for a limited time it collimates the secondary wind towards the polar directions.
The higher density in this collimated wind further enhances the intensity of
the He~II~$\lambda 4686$ line by a large factor (Kunasz 1980), leading to the peak
in the He II $\lambda 4686$ line intensity.
This collimated wind could also be the low-energy analog of the X-ray jet
(Behar et al. 2006).
After the short peak, the secondary wind is totally shut down (Soker 2005b; Akashi et al.\ 2006),
and hence the He~II~$\lambda 4686$ line as well as the X-rays are shut down as well.
After periastron when the secondary wind is resurrected (and the X-rays return) the secondary wind is
back to its initial pre-accretion form, where the He~II line does not exist.
{{{ In this phase we expect to see only the blue-shifted (approaching) side of the
collimated wind.
The reason is that the inclination of \etc~ is $i=42^\circ$ (i.e., the orbital plane is
tilted by $48^\circ$ from an edge-on view; Smith 2002), and the He~II~$\lambda 4686$
line is formed very close to the stellar surface, $r_2 \sim 1.5 R_2$.
Therefore, the star itself will block most of the He~II~$\lambda 4686$ red-shifted line,
and the rest will be blocked by the dense region at the base of the wind at
$R_2<r_2 \la 1.2 R_2$. }}}

The sharp decrease of the the He~II~$\lambda 4686$ line by a factor of $\sim 5-10$
over about 10 days is accounted for by (1) The quick dying out of the secondary wind;
and (2) the speed of the secondary wind entering the shock is much lower.
The lower speed is due to the X-ray ionization and heating and also
due to the stagnation region moves closer into the acceleration zone.
The X-ray radiation will therefore be softer, hence absorbed more,
hence a lesser affect on the acceleration zone.

We note that M2006 briefly mention that the line might be formed by the winds of the two
stars; they mainly consider the primary wind.
However, M2006 assume that the He~II emission originates
from photoionized regions near the X-ray shock fronts.
The problem with this assumption, as noted by M2006, is that the He~II~$\lambda 4686$
line peak occurs well after the X-ray emission has started to drop.
{{{ SD04 also considered this possibility, but they also considered an alternative,
where the He~II~$\lambda 4686$ originates in the acceleration zone of the secondary wind.
We agree with this second suggestion of SD04. }}}

{{{{ We end this subsection by emphasizing again three points of our model.
(1) The He~II~$\lambda 4686$  is formed in the acceleration zone of the secondary wind.
The sensitivity of this line to wind properties (Kunasz 1980; Venero et al. 2002) implies
that changes in only a fraction of the acceleration zone can lead to a large variation
in the line intensity.
(2) The primary star has no direct
role in the changes in the He~II line properties; its only role is in supplying the gas
that collides with the secondary wind, leading to X-ray emission, and in supplying
the accreted gas during the accretion phase.
(3) The energy in the He~II line comes from the wind and the secondary star.
Both the kinetic power in the secondary wind and the secondary luminosity are
much larger than the maximum He~II line luminosity.
Therefore, there is no problem in supplying the energy to this line. }}}}

\subsection{Specific Angular Momentum}
We show now that the gas in the stagnation point has indeed enough
specific angular momentum to be accreted from the equatorial plane,
which could form a collimated wind, or even two opposite jets.
We take the shocked primary wind near the stagnation point to move slowly
relative to the stagnation point, hence it is corotating with the binary orbit
at an angular velocity $\omega_b=v_{\rm peri} r_{\rm peri} /r^2$,
where $v_{\rm peri}$ is the binary relative velocity at periastron and
$r_{\rm peri}$ is the orbital separation at periastron.
The specific angular momentum of the matter near the stagnation point
around the secondary is $j_{s1}=D_{g2}^2 \omega_b$.
For the relevant orbital phases we see from Figure 1 that $r/D_{g2} \simeq 4$,
from which we find $j_{s1} \simeq 0.06 v_{\rm peri} r_{\rm peri}$.
To check the degree by which the accretion onto the secondary depart from sphericity,
and whether an accretion disk might form, $j_{s1}$ should be compared with the specific
angular momentum of a test particle performing Keplerian motion on the secondary
equator $j_2 =  (G M_2 R_2)^{1/2}$.
Substituting $v_{\rm peri} \simeq 390 \km \s^{-1}$,
$r_{\rm peri} = 1.66 \AU$ (Figure 1), and $M_2=30 M_\odot$ we find
\begin{equation}
\frac{j_{s1}}{j_2} \simeq 0.8
\left( \frac{R_2}{20 R_\odot} \right)^{-1/2}.
\label{j21}
\end{equation}
This shows that the collapsing material at the beginning of the accretion
phase (and only at the beginning, before the Bondi-Hoyle type accretion takes place)
has large enough specific angular momentum such that most of the gas is
accreted from near the equator.
This will force the secondary wind to flow along the poles, hence having a
much higher density at the wind base.
Such a high density might substantially increase the He~II~$\lambda 4686$
line luminosity (Kunasz 1980).
Shortly after the collapse stage, the secondary wind is shut down, reducing to zero the
He~II~$\lambda 4686$ line and the X-ray emission.
Formation of an accretion disk is marginal, and more accurate calculations are required
to check whether an accretion disk can be formed, and whether two opposite jets
might be launched.
This ends our explanation to the He II $\lambda 4686$ emission peak.

% ==========================================================
\section{DISCUSSION AND SUMMARY}
\label{sec:event}
% ==========================================================

In the present paper we continue to explore the different aspects of the
winds interaction in the massive binary system \etc\, and study the
anomalously high He~II $\lambda$4686 line appearing just before
periastron and quickly disappearing immediately thereafter.
Based on the He~II $\lambda$4686 line emission from O-stars and their
modelling in the literature, we postulate that the
He~II $\lambda$4686 line comes from the acceleration zone of the
secondary stellar wind.
The large increase in the He~II $\lambda$4686 line intensity is attributed
to small changes in the properties of the secondary stellar wind.
We suggest that the formation and acceleration of the secondary stellar wind is
affected by the winds interaction in two ways.

The first effect is due to the enhanced X-rays emitted by the shocked secondary wind.
The secondary stellar wind is shocked as it encounter the primary stellar wind.
The X-ray emission could reduce the acceleration of the secondary stellar
wind (section 2.3.1; SK90), and hence increase its density in the
acceleration zone, which also heats the wind.
Under the appropriate conditions, which we suggest exist in the acceleration
zone of the secondary wind in \etc,  higher density (Kunasz 1980)
and/or higher temperature (Venero et al. 2002) increase the intensity of the
He~II~$\lambda 4686$ line emission.

The second effect is due to the accreted primary stellar wind, which
affects the geometry of the secondary wind.
Close to periastron passage, the secondary star starts to accrete from the
shocked primary stellar wind (Soker 2005b; Akashi et al. 2006).
Because of the high specific angular momentum of this accreted gas (eq. \ref{j21}),
it will be accreted from near the equatorial plane.
The equatorial accreted mass will collimate the secondary wind along the polar
directions, and might even form an accretion disk and launch two opposite transient jets,
as suggested recently by X-ray observations (Behar et al. 2006).
The polar outflow from the secondary will have much higher densities than the isotropic wind,
and it is, therefore, expected to be a much stronger He~II~$\lambda 4686$ emitter.

Our explanation for the temporal evolution of the He~II~$\lambda 4686$ line is given
in section 3.2.
The presently proposed explanation for the He~II $\lambda$4686 high intensity
disputes previous ones (SD04; M2006), which entailed photoionization of He exclusively
by the X-rays.
{{{ We agree, on the other hand, with the other suggestion of SD04 that the
He~II~$\lambda 4686$ line originates in the acceleration zone of the secondary wind,
and that the slow rise to maximum occurs on the side facing the winds interaction zone. }}}
The high observed intensity of over 10$^{53}$ line photons
emitted over a period of about 20 days, and while the X-ray flux is sharply dropping,
imply a prohibitively high photoionizing X-ray flux, which the present model
does not require.

Finally, we note that the influence of the X-ray emission on the secondary stellar
wind, in making its velocity lower, will cause the stagnation point to
be closer to the secondary star compared with assumptions made in earlier papers
in the series (Soker 2005b; Akashi et al. 2006). Although this effect is not large,
it reinforces accretion of primary shocked wind material near periastron,
the key element in the accretion model.

We thank an anonymous referee for useful comments.
This research was supported by a grant (\#28/03)
from the Israel Science Foundation and by a grant from the
Asher Space Research Institute at the Technion.

%====================================================================
%   FIGURE 1
\bigskip
%====================================================================
%% \begin{figure}
%% \epsfysize=8cm
%% \epsffile{zt1-new.eps}
%% \label{fig1}
%% \caption{Aaaaaaaaaaaaa}
%% \end{figure}
%====================================================================
%====================================================================
%====================================================================
%   FIGURE 2
%====================================================================
%====================================================================
%====================================================================
%\bigskip
%   FIGURE 3
%====================================================================

\end{document}